\documentclass[%
 aip,
 amsmath,amssymb,
 reprint,%
]{revtex4-1}
\usepackage{graphicx}  
\usepackage{dcolumn}   
\usepackage{bm}        
\usepackage{amssymb}   
\usepackage{amsmath}   
\usepackage{braket}    
\usepackage{mathtools}
\usepackage[normalem]{ulem}
\usepackage{fancyhdr}
\usepackage[hidelinks]{hyperref} 
\usepackage{cleveref}
\usepackage{float}
\usepackage{xcolor}
\usepackage[caption=false]{subfig} 
\usepackage{ragged2e} 
\DeclareCaptionJustification{justified}{\justifying}

\begin{document}

\title{Engineering superpositions of N00N states using an 
	asymmetric \\ non-linear Mach-Zehnder interferometer}
\author{Richard J. Birrittella$^{1,2}$, Paul M. Alsing$^{1}$,  James Schneeloch$^{1}$, \\Christopher C. Gerry$^{3}$, Jihane Mimih$^{4}$ and Peter L. Knight$^{5}$ \\
\textit{\textit{$^{1}$Air Force Research Laboratory, Information Directorate, Rome, NY, USA, 13441}
	\\
\textit{$^{2}$Griffiss Institute, 592 Hangar Rd STE 200, Rome, NY 13441}\\	
\textit{$^{3}$Department of Physics and Astronomy, Lehman College,\\
The City University of New York, Bronx, New York, 10468-1589,USA} 
}\\
\textit{$^{4}$Department of Electrical and Computer Engineering, Naval Postgraduate School, 
\\1 University Circle, Monterey, California 93943, USA}\\
\textit{$^{5}$Blackett Laboratory, Imperial College, London SW72AZ, UK  } 
}

\date{\today}


\begin{abstract}
We revisit a method for mapping arbitrary single-mode pure states into superpositions of N00N states using an asymmetric non-linear Mach-Zehnder interferometer (ANLMZI). This method would allow for one to tailor-make superpositions of N00N states where each axis of the two-mode joint-photon number distribution is weighted by the statistics of any single-mode pure state. The non-linearity of the ANLMZI comes in the form of a $\chi^{\left(3\right)}$ self-Kerr interaction occurring on one of the intermediary modes of the interferometer. Motivated by the non-classical interference effects that occur at a beam splitter, we introduce inverse-engineering techniques aimed towards extrapolating optimal transformations for generating N00N state superpositions. These techniques are general enough so as to be employed to probe the means of generating states of any desired quantum properties.  
\end{abstract}


\maketitle

\section{\label{sec:intro} Introduction}

\noindent Jon Dowling had a talent for coming up with memorable names and phrases, and probably the most enduring example of this is his coining the words “$N00N$ states,” with the insistence that it not be written as “NOON states.” He recognized early on that two-mode field states possessing bimodal joint photon number probability distributions that were widely separated in the number
states basis, as is the case for $N00N$ states, were key to attaining Heisenberg-limited sensitivities in quantum optical interferometry. $N00N$ states themselves are notoriously difficult to generate, but continuous variable superpositions of $N00N$ states are more easily generated. 
In honor and memory for Jon's pioneering contributions in this field, 
these investigations on the notion of N00N states are the subject of this paper. \\

\noindent Quantum mechanical states of light have been studied extensively in the field of quantum metrology \cite{ref:Caves,ref:Dowling_N00N,ref:Birrittella_PSsvs,ref:Rafsanjani,ref:Wang,ref:Birrittella_ParityReview,ref:Chen}, where one is interested in performing highly resolved and sensitive measurements of signals like, for example, what one would expect to detect from gravitational waves \cite{ref:Aasi_LIGO} (also see Barsotti \textit{et al.} \cite{ref:Barsotti} and references therein) or for the precise measurement of transition frequencies in atomic (ion) spectroscopy \cite{ref:Bollinger}.  The advantage one gains over using classical fields is the ability to exploit inherently quantum characteristics of the state such as entanglement, squeezing or some other non-classical property \cite{ref:Pathak_ClassVSNonclass}.  The goal lies in reaching the greatest degree of phase-measurement sensitivity afforded by quantum mechanical states (for linear phase shifts): the Heisenberg limit (HL).  The HL serves as an improvement over the standard quantum limit (SQL) of phase sensitivity, which represents the best sensitivity attainable by classical and classical-like states,  by a factor of the SQL itself, i.e.

\begin{equation}
	\Delta\phi_{\text{HL}}= \frac{1}{\bar{n}} =\left(\Delta\phi_{\text{SQL}}\right)^2,
	\label{eqn:inter_1}
\end{equation}

\noindent where $\bar{n}$ is the (conserved) average photon number in the system. This limit can be understood from the heuristic relation $\Delta\phi\Delta n \simeq 1$ by considering the much-discussed $N00N$ states \cite{ref:Dowling_N00N} of the general form

\begin{equation}
	\ket{\psi_{N00N}}=\frac{1}{\sqrt{2}}\big(\ket{N,0}_{a,b}+e^{i\Phi_N}\ket{0,N}_{a,b}\big)
	\label{eqn:intro_1}
\end{equation} 

 \noindent For this case the uncertainty in photon number is equal to the total photon number itself $N$ making the phase uncertainty $\Delta\phi\simeq 1/N$. Superpositions of such states cannot be made through typical beam splitters but rather have been demonstrated to require some form of nonlinear interaction \cite{ref:Sanders_ECS,ref:Gerry_ANLI}.  Such superpositions have also been discussed in relation to Heisenberg-limited interferometry. However, the state alone is just one ingredient in the interferometric scheme.  The other is in choosing an optimal detection observable.  For example, it has been shown for entangled coherent states (ECS) of the form $\ket{\alpha,0}+\ket{0,\beta}$, where $|\alpha|=|\beta|$, that one obtains the HL for parity-measurement-based interferometry. In fact, parity-based measurements are the realization of an earlier Hermitian operator of the form
 
 \begin{equation}
     \hat{\Sigma}_N=\ket{N,0}_{a,b}\bra{0,N}+\ket{0,N}_{a,b}\bra{N,0},
     \label{eqn:Intro_Added_1}
 \end{equation}
 
 \noindent whose expectation value displays interference fringes that oscillate with frequency $N\phi$.  This operator has been shown to yield the HL for the case of $N00N$ states, where intensity-difference measurements fail to capture a phase-dependent measurement \cite{ref:Dowling_N00NProj1,Dowling_N00NProj2}.  It turns out that parity-based measurements yield the minimum phase uncertainty, saturating the quantum Cram\'{e}r-Rao bound \cite{ref:Braunstein_QFI}, for all path-symmetric input states \cite{ref:Hoffman_PathSym,ref:Dowling_PathSym}, making it the optimal detection observable for most interferometric experiments. Parity detection has also been shown to perform at the HL for quantum metrology using an SU(1,1) interferometer, characterized by replacing the beam splitters with down-converters operating under the parametric approximation \cite{ref:Dong_SU11}.\\

\noindent In this paper, we revisit the so-classed asymmetric non-linear Mach-Zehnder interferometer (ANLMZI), characterized by one intermediary mode of the interferometer passing through a $\chi^{\left(3\right)}$ self-Kerr medium. We aim to show how one can generate arbitrary $N00N$ state superpositions weighted by the statistics of any single-mode pure-state using such a device.  The resulting transformation can be viewed as the mapping in $N00N$ state space

\begin{equation}
	\ket{\psi}_a\otimes\ket{0}_b \stackrel{\text{ANLMZI}}{\longrightarrow}\frac{e^{i\tfrac{\pi}{4}}}{\sqrt{2}}\left(\ket{\psi,0}_{a,b}-i\ket{0,\psi}_{a,b}\right).
\end{equation}

\noindent We go on to investigate the validity of the case in which one has a cross-Kerr medium in lieu of a self-Kerr medium and show that it proves a viable means of generating $N00N$ states, where one need only perform an $N\pi/2$-phase shift prior to the second beam splitter of the interferometer. We also explore a more general means of generating $N00N$ state superpositions through inverse-engineering from a presupposed form of output state from a symmetric beam splitter. Although it is discussed in this particular context, the inverse-engineering techniques involved can be generalized to probe the validity of generating any state with the desired properties.  We find that such mappings are not generally unitary, but could potentially be realized experimentally via boson-mode operations and state-reductive projections. \\

\noindent The paper is organized as follows: In Section \ref{sec:interferometry} we briefly review some relevant works on obtaining Heisenberg-limited phase sensitivity in quantum optical interferometry and discuss some interesting non-classical interference effects that occur at a beam splitter, which we term the \textit{extended Hong-Ou-Mandel} (eHOM) effect.  In Section \ref{sec:mapping} we investigate utilizing eHOM-like interference effects by means of the inverse-engineering of the output joint-photon probability distributions of a balanced beam splitter to create arbitrary superpositions of N00N states, and demonstrate one potential experimental realization using an asymmetric non-linear Mach-Zehnder interferometer.  We close in Section \ref{sec:closing} with a discussion of our findings and some concluding remarks.  For completeness, we also include brief supplementary material reviewing the Schwinger realization of the SU(2) Lie algebra in Appendix \ref{app:secA} as well as the corresponding Wigner-\textit{d} rotation elements in Appendix \ref{app:secB}.  

\begin{figure*}
	\subfloat[][]{\includegraphics[width=0.33\linewidth,keepaspectratio]{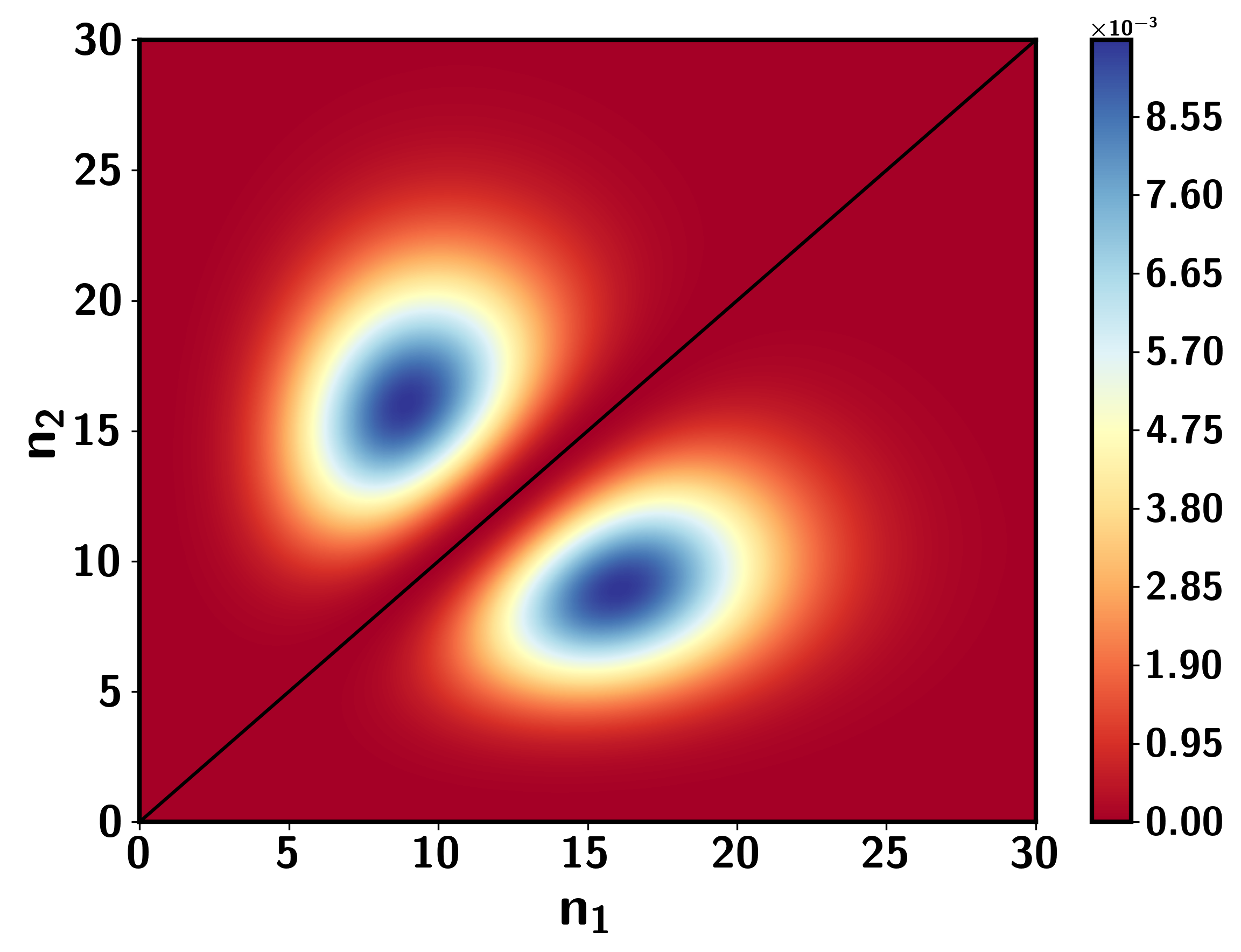}
		\label{fig:coh_n_1}}
	\subfloat[][]{\includegraphics[width=0.33\linewidth,keepaspectratio]{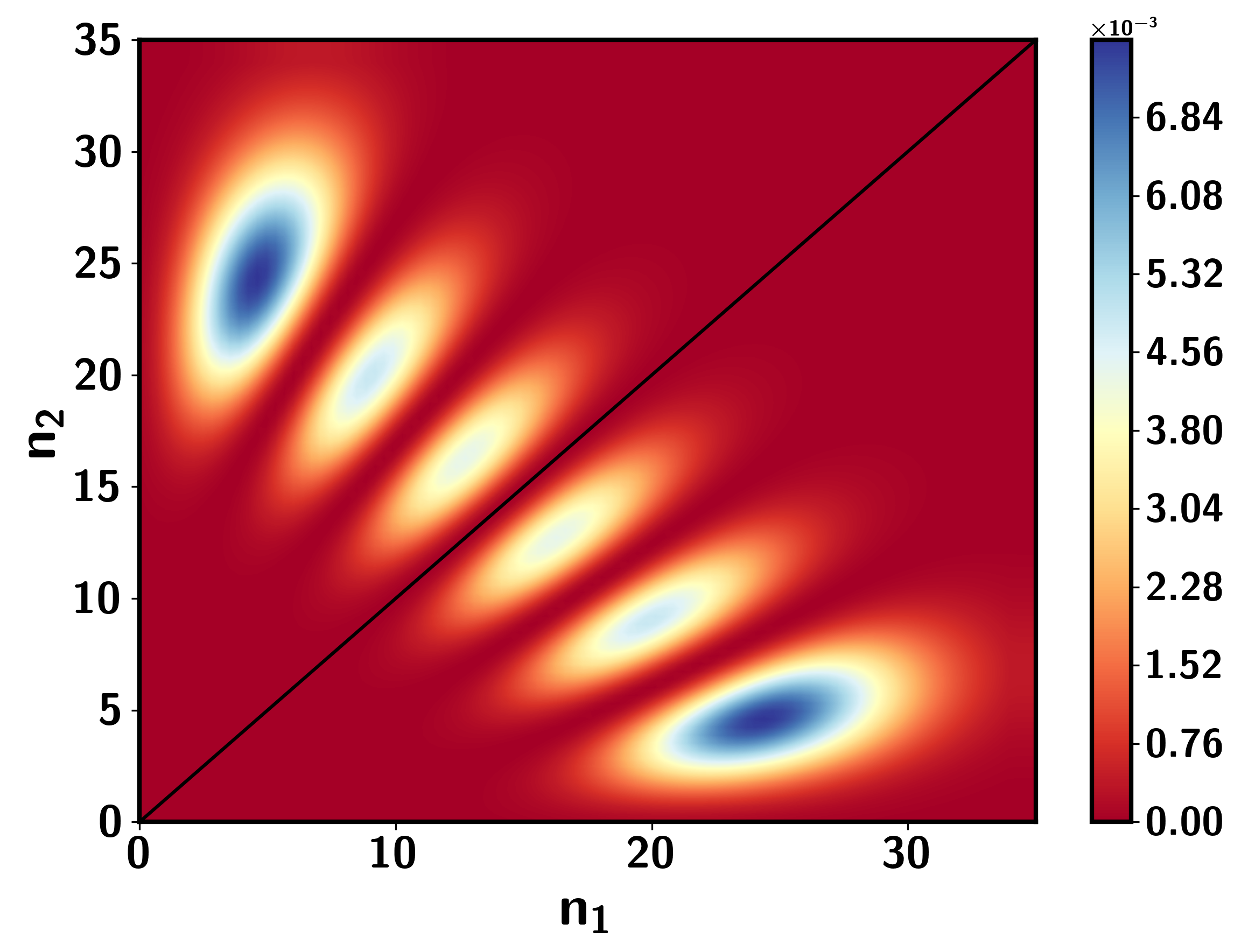}
		\label{fig:coh_n_5}}
	\subfloat[][]{\includegraphics[width=0.33\linewidth,keepaspectratio]{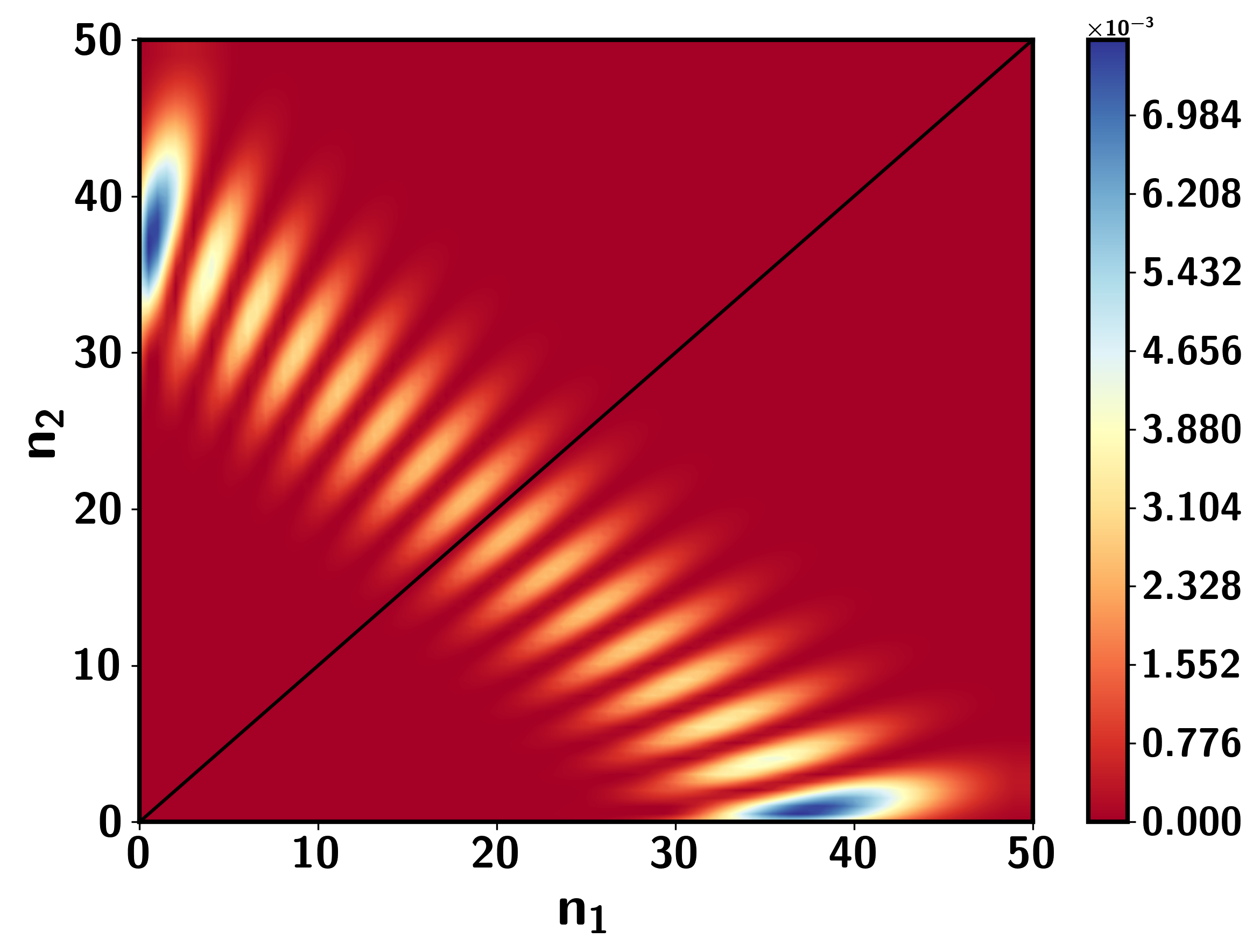}
		\label{fig:coh_n_15}}
	\caption{Countour plots of the (interpolated) joint-photon number distribution after beam splitting for the case of mixing coherent light with amplitude $\alpha=5$ with a Fock state \protect\subref{fig:coh_n_1} $n=1$, \protect\subref{fig:coh_n_5} $n=5$, and \protect\subref{fig:coh_n_15} $n=15$.  The Fock state initially occupying the $b$-mode can be deduced by taking $P-1$ where $P$ is the number of peaks in the distribution. For increasing $n$, the largest peaks of the distribution are being `pushed' towards the axes, reminiscent of a N00N superposition state. Note the black line denotes the CNL: destructive interference of all $\ket{n',n'}_{a,b}$ correlated photon states.} \label{fig:Coherent_Fock_Contours}
\end{figure*}

\section{\label{sec:interferometry} Interference at a beam splitter: Towards Heisenberg-limited Interferometry}

\noindent  In light of the preceding discussion, much work has gone into generating states that display the largest separation in their joint-photon number distribution prior to the second beam splitter of the MZI, as they, much like the $N00N$ states themselves, tend towards offering the greatest phase sensitivity (i.e. the smallest phase uncertainty). Analogous to this, one is interested in generating states with a large degree of path-entanglement \cite{ref:Kok_PathEntang,ref:Lee_PathEntang}.  Many schemes involve leveraging the well known result that coherent light mixed with an even or odd Schr\"{o}dinger cat at a beam splitter produces coherent superpositions of N00N states \cite{ref:Sanders_ECS,ref:Gerry_ANLI} by, for example, replacing the cat state with a photon-added \cite{ref:Wang} or photon-subtracted \cite{ref:Birrittella_PSsvs, ref:Leuchs} single-mode squeezed vacuum states. Such schemes would conditionally generate the cat-like states, resulting in a two-mode distribution similar to that of a true $N00N$ state superposition. Others \cite{ref:Afek_Corner,ref:Birrittella_Interference} have pointed out that when beam splitting coherent light mixed with a squeezed vacuum, where the states are of equal low intensities, one obtains a joint-photon number distribution peaked at the vacuum and with a thermal-like distribution along each of the axes: the state is coined a `corner state' by the original authors.  \\

\noindent Another such case would be mixing coherent light with a Fock state of discrete photon number $N$, such that the input state to the MZI is given by $\ket{\psi_\text{in}}=\ket{\alpha}_a\ket{N}_b$. For constant coherent amplitude, this state displays a joint-photon number distribution after the first beam splitter that becomes increasingly localized towards the axes for increasing values of $N$.  This can be seen in Fig.~\ref{fig:Coherent_Fock_Contours} for $N=1,5,15$ where the peaks are successively migrating away from the central (diagonal black) line towards the $n_1=0$ and $n_2=0$ axes. These states were investigated for use in quantum optical interferometry by Birrittella \textit{et al.} \cite{ref:Birrittella_Interference} who found the minimum phase uncertainty to be 

\begin{equation}
	\Delta\phi^{\left(\alpha,N\right)}_{\text{min}} = \frac{1}{\sqrt{|\alpha|^2+N\left(1+2|\alpha|^2\right)}},
	\label{eqn:inter_2}
\end{equation} 

\noindent which in the limit of $|\alpha|^2=N=\bar{n}_{\text{total}}/2\gg 1$ becomes $\Delta\phi^{\left(\alpha,N\right)}_{\text{min}}\to \sqrt{2}/\bar{n}_{\text{total}}$, proportional the the HL. In practice, however, generation of large-photon-number Fock states is experimentally impractical. Instead, other means should be considered to achieve phase sensitivity that approaches the HL. \\

\noindent As was first noticed by Birrittella \textit{et al.} \cite{ref:Birrittella_Interference} and further studied by Alsing \textit{et al.} \cite{ref:Alsing_eHOM}, the parity of a non-classical state has a profound effect on the output state statistics when mixed at a $50:50$ beam splitter with any other state.  More specifically, if one of the input ports is occupied by an odd Fock state such that $\hat{\Pi}\ket{2k+1}=-\ket{2k+1}$, where $k\in \mathbb{Z}^{0+}$ and where $\hat{\Pi}=\left(-1\right)^{\hat{n}}$ is the parity operator, then the resulting diagonal probabilities of the joint-photon number distribution can be written as 

\begin{equation}
	P\left(n,n|2k+1\right) \propto \left|\cos\theta\times\text{poly}_{k}\left(n,\theta\right)\right|^2,
	\label{eqn:inter_3}
\end{equation}     

\noindent where the beam splitter is defined such that the transmittance is given by $T=\cos^2\tfrac{\theta}{2}$ and where $\text{poly}_{k}\left(n,\theta\right)$ is an arbitrary polynomial function in $n,\theta$ of order $k$. It is clear from Eq.~\ref{eqn:inter_3} that for a $50:50$ beam splitter, all correlated photon-number states $\ket{n,n}$ of the output state will destructively interfere, resulting in a line of contiguous zeros in the output probability distribution known as the central nodal line (CNL). This effect can be observed in the joint-photon number distribution contours of Fig.~\ref{fig:Coherent_Fock_Contours} for the case of mixing coherent light with an odd Fock state at a balanced beam splitter.  We can further illustrate this with the simplest case where we start with the $\ket{1}_b$ photon state: mixing this with the smallest Fock state such that the output can contain a correlated state (i.e. an even number of total photons), the $\ket{1}_a$ photon state, results in the well-known Hong-Ou-Mandel (HOM) effect for which destructive interference eliminates the $\ket{1,1}_{a,b}$ output. From this, one can consider the more general case of mixing Fock states: $\ket{2k+1,1}_{a,b}$, for which the resulting distribution will not contain the state $\ket{k+1,k+1}_{a,b}$ state. One can then infer that this effect will hold true for any superposition state of definite odd parity $\hat{\Pi}\ket{\psi}=-\ket{\psi}$, such as, and for example, odd cat states as well as photon subtracted/added squeezed vacuum states. Provided one port of a $50:50$ beam splitter is occupied by an eigenstate of $\hat{\Pi}$ with eigenvalue $\left(-1\right)$, the resulting probability distribution will contain a CNL. It is for this reason, that the authors of \cite{ref:Alsing_eHOM} colloquially refer to this as the extended Hong-Ou-Mandel effect (eHOM), for which the HOM effect (i.e. the destructive interference of the $\ket{1,1}$ state at a balanced beam splitter) is a limiting case.  Another consequence of the non-classicality of the input state is in the off-diagonal lines of destructive interference which can be seen in Figs.~\ref{fig:coh_n_5} and Fig.~\ref{fig:coh_n_15}, which can be viewed as a form of interference fringes for the joint-photon distribution; these are referred to in Ref.\cite{ref:Gerry_NonDemoParity} as pseudo-nodal curves (PNCs), and will occur whenever one port of a beam splitter contains a state of definite even or odd parity. These PNCs do not constitute lines of perfect zeros but rather serve as local minima for the distribution, effectively carving out valleys in the distribution. For example, for the $\ket{N,\alpha}_{a,b}$ input, one can expect to find $N$ valleys and $N+1$ peaks, which can be verified from Fig.~\ref{fig:Coherent_Fock_Contours}.  For a more detailed discussion on the topic of the eHOM effect, see Alsing \textit{et al.} \cite{ref:Alsing_eHOM}.     

\section{\label{sec:mapping} Mapping single-mode states to superpositions of N00N states}

\noindent For the case of the eHOM effect discussed above, the quantum amplitude interference caused by the balanced beam splitter carves out valleys in the output joint-photon number distribution. As we have seen in Fig.~\ref{fig:Coherent_Fock_Contours}, Fock states of larger photon number interfering with a coherent state on a balanced beam splitter pushes the output probability distribution towards the axes (edges). In this section we introduce inverse-engineering techniques aimed towards exploring the possibility of turning these interference valleys in the distribution into interference basins.

\subsection{\label{subsec:ANLMZI} The asymmetric non-linear Mach-Zehnder interferometer}

\begin{figure}
	\includegraphics[width=0.95\linewidth,keepaspectratio]{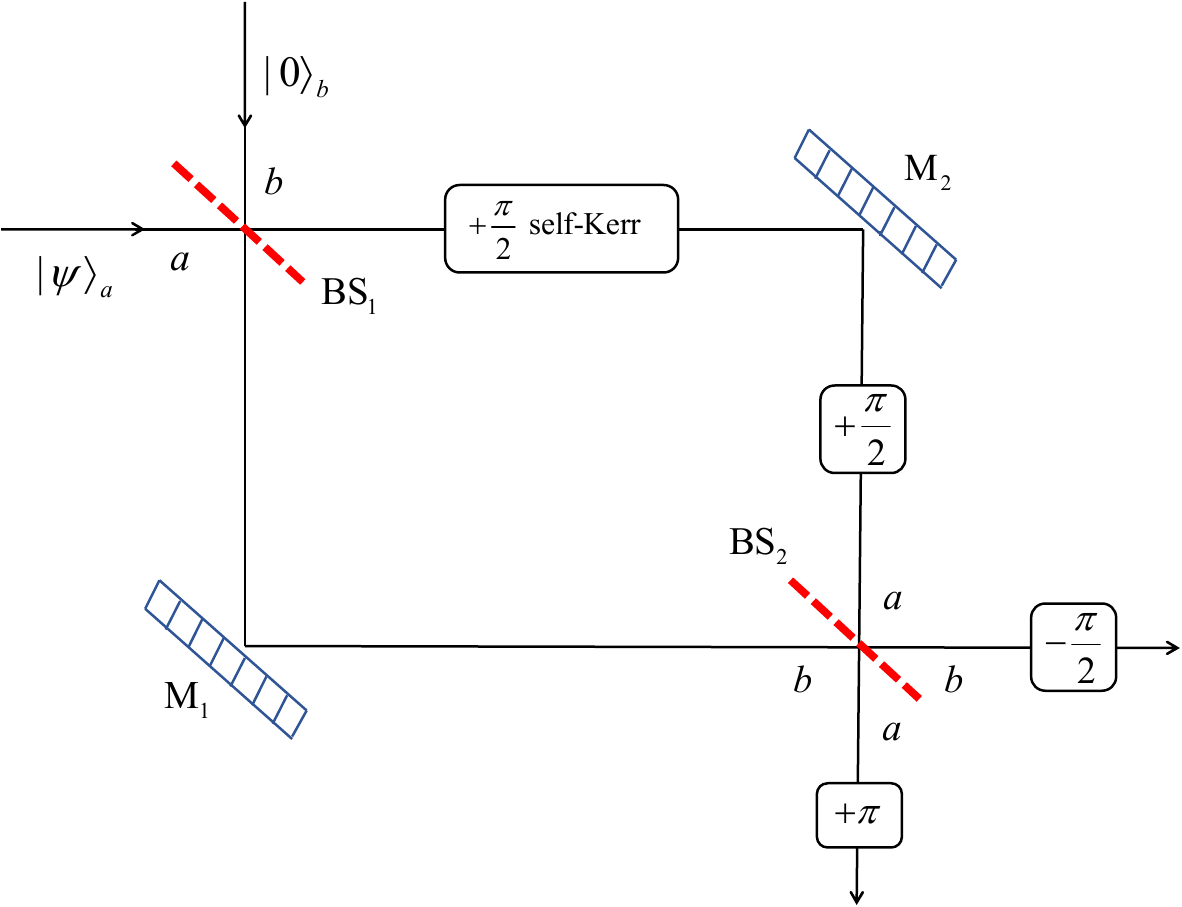}
\caption{A sketch of the set-up.  The assymetric nonlinear MZI is characterized by a self-Kerr interaction on the intermediary $a$-mode prior to the second beam splitter.  The other boxes along the beam paths represent linear phase-shifters, which for a phase $\varphi$ are expressed as $e^{i\varphi\hat{n}_{a\left(b\right)}}$, for the $a$- and $b$-modes, respectively.}
	\label{fig:sketch}
\end{figure}

\noindent We begin by revisiting the ANLMZI, for which a schematic is given in Fig.~\ref{fig:sketch}. It is comprised of a standard Mach-Zehnder interferometer constructed with two 50:50 beam splitters described by the transformation \cite{ref:Campos_SU2,ref:Yurke_SU2}

\begin{equation}
	\hat{\mathcal{U}}_{\text{BS}}^{\left(a,b\right)} = e^{i\tfrac{\pi}{4}\left(\hat{a}^\dagger\hat{b}+\hat{a}\hat{b}^\dagger\right)}=e^{i\tfrac{\pi}{2}\hat{J}_x.},
	\label{eqn:campos_1}
\end{equation} 

\noindent where $\big(\hat{a},\hat{b}\big)$ are the $a$- and $b$-mode boson operators and where in the last step we express these operators in terms of the Schwinger realization of SU(2) (see Appendix \ref{app:secA} for more detail). Note that the transformation is defined in such as a way as to introduce a phase factor of $i$ in the reflected mode. The path-length difference between arms of the interferometer is realized as a phase-shift occurring in one arm and is described by the transformation $\hat{\mathcal{U}}^{\left(a\right)}_{\text{PS}}\left(\phi\right)=e^{i\phi\hat{n}_a}$, where $\hat{n}_a$ is the $a$-mode number operator with the action $\hat{n}_a\ket{n}_a=n\ket{n}_a$. The non-linearity arises through the self-Kerr interaction on the intermediary $a$-mode described by the interaction Hamiltonian \cite{ref:Kitagawa}

\begin{equation}
	\hat{H}^{\left(a\right)}_{\text{self-Kerr}} = \hbar\chi\hat{a}^{\dagger\;2}\hat{a}^2=\hbar\chi\left(\hat{n}_a^2-\hat{n}_a\right),
	\label{eqn:campos_2}
\end{equation}    

\noindent where $\chi$ is proportional to the third-order non-linear susceptibility $\chi^{\left(3\right)}$ of the medium.  Note that many authors adopt a form of the self-Kerr interaction in which the linear phase term is omitted \cite{ref:Sanders_ECS,ref:Milburn}.  This linear phase can be easily compensated for in the ANLMZI through the use a linear phase-shifter.  For this reason we will keep this term in the analysis that follows.  The unitary transformation associated with the self-Kerr interaction is then given by

\begin{align}
	\hat{\mathcal{U}}^{\left(a\right)}_{\text{self-Kerr}}\left(\kappa\right) &= e^{-i\frac{t}{\hbar}\hat{H}^{\left(a\right)}_{\text{self-Kerr}}} = e^{-i\kappa\left(\hat{n}_a^2-\hat{n}_a\right)}, \nonumber \\
	&= e^{-i\kappa\hat{n}_{a}^2}e^{i\kappa\hat{n}_a},
	\label{eqn:campos_3}
\end{align}  

\noindent where $t=l/v$ is the time light takes to propagate through the non-linear medium, $l$ is the length of the medium and $v$ is the velocity of light in the medium. Further, we define the scaled time $\kappa=\chi t$.  \\

\noindent In our scheme, we assume a sufficiently large degree of non-linearity (interaction time or length-of-medium) such that $\kappa=\pi/2$. We point out that many proposals in the literature exploiting the use of third-order non-linearities in the form of self- or cross-Kerr interactions rely on this assumption \cite{ref:Chaba,ref:Imoto,ref:Haus,ref:Gerry_ANLI,ref:Gerry_NonDemoParity}. 

\noindent A scheme similar to Fig.~\ref{fig:sketch} was used by Gerry \textit{et al.} \cite{ref:Gerry_ANLI} (in their description, the intermediary phase shift preceded the Kerr interaction; this ordering does not impact the final state) to show how one can generate maximally ECS.  Starting  with an input state of the form $\ket{\psi_\text{in}}=\ket{\alpha,0}_{a,b}$, where $\ket{\alpha}\propto\sum_{n}\alpha^n/\sqrt{n!}\ket{n}$ is the usual coherent state, they arrived at the output state 

\begin{equation}
	\ket{\psi_\text{out}} = \frac{e^{i\tfrac{\pi}{4}}}{\sqrt{2}}\left(\ket{\alpha,0}_{a,b} - i\ket{0,\alpha}\right),
	\label{eqn:campos_4}
\end{equation}

\noindent where in their derivation they took advantage of a result first pointed out by Yurke \textit{et al.} \cite{ref:Yurke_YScat} for unitarily generating cat states via a self-Kerr non-linear interaction

\begin{equation}
	e^{-i\frac{\pi}{2}\hat{n}^2}\ket{\beta} = \frac{e^{-i\frac{\pi}{4}}}{\sqrt{2}}\left(\ket{\beta}-i\ket{-\beta}\right).
	\label{campos_5}
\end{equation}

\noindent Eq.~\ref{eqn:campos_4} informs us that one can generate superpositions of $N00N$ states weighted by the coefficients of a coherent state.  \\

\noindent Building upon these results, we will endeavor to show this scheme will work for \textit{any} single-mode state.  Once again following the schematic shown in Fig.~\ref{fig:sketch}, for a general $N$-photon Fock state initially occupying the $a$-mode, we start with the total input state

\begin{equation}
	\ket{\psi_\text{in}} = \ket{N}_a\otimes\ket{0}_b=\ket{N,0}_{a,b} = \ket{j,j},
	\label{eqn:derivation_1}
\end{equation}

\noindent where in the last step, we have utilized the Schwinger realization of the SU(2) Lie algebra (see Appendix \ref{app:secA} to express two boson modes in terms of an `angular momentum' state (i.e. a multiplet state of \textit{su}(2)). The state after the first beam splitter is then given by

\begin{align}
	\ket{\psi_\text{BS1}} &=\sum_{n=0}^{N} i^{N-n}\;d_{n-\tfrac{N}{2},\tfrac{N}{2}}^{N/2}\left(\frac{\pi}{2}\right)\ket{n,N-n}_{a,b}, \nonumber\\
	&= \sum_{n=0}^{N}C_{n}^{\left(N\right)}\ket{n,N-n}_{a,b},
	\label{eqn:derivation_2}
\end{align}

\noindent where $d_{m',m}^{j}\left(\beta\right)$ represent the Wigner-$d$ rotation matrix elements briefly discussed in Appendix \ref{app:secB} and where in the last line of Eq.~\ref{eqn:derivation_2} we have consolidated terms into the probability amplitudes $C_{n}^{\left(N\right)}$. Using the transformations for the self-Kerr interaction as well as a linear phase-shift on the $a$-mode, the state prior to the second beam splitter is given by

\begin{equation}
	\ket{\psi_\text{pre-BS2}} =  \sum_{n=0}^{N}C_{n}^{\left(N\right)}\left(-i\right)^{n^2}\ket{n,N-n}_{a,b}.
	\label{eqn:derivation_3}
\end{equation}

\noindent Finally, the state after the second beam splitter and subsequent phase-shifters can be written as 

\begin{equation}
	\ket{\psi_\text{out}} = \sum_{n'=0}^{N}\gamma_{n'}^{\left(N\right)}\ket{n',N-n'}_{a,b},
	\label{eqn:derivation_4}
\end{equation}

\noindent where the new probability amplitudes can be simplified to 

\begin{align}
	\gamma_{n'}^{\left(N\right)} &= \left(-1\right)^{n'}\sum_{n=0}^{N}\left(-i\right)^{n^2}d_{n'-\tfrac{N}{2},n-\tfrac{N}{2}}^{N/2}\left(\frac{\pi}{2}\right)\times\nonumber\\
	&\;\;\;\;\;\;\;\;\;\;\;\;\;\;\;\;\;\;\;\;\;\;\;\;\;\;\;\;\;\;\;\;\;\;\;\;\;\times d_{n-\tfrac{N}{2},\tfrac{N}{2}}^{N/2}\left(\frac{\pi}{2}\right).
	\label{eqn:derivation_5}
\end{align}

\noindent It can then be shown \footnote{In practice, the Wigner-$d$ elements are expressed in terms of Hypergeometric functions that are not easily simplified. Consequently, the derivation of the state coefficients can be confirmed numerically. The important point to note is that one only gets the necessary cancellations of the off-axis probabilities due to the non-linear phase shift $(-i)^{Na^2}$.} that the state coefficients $\gamma_{n'}^{\left(N\right)}$ are given by 

\begin{equation}
	\gamma_{n'}^{\left(N\right)} = 
	\begin{cases} 
		\frac{1}{\sqrt{2}}e^{i\frac{\pi}{4}} & n'=N,\\
		-\frac{i}{\sqrt{2}}e^{i\frac{\pi}{4}} & n'=0, \\
		0 & \text{otherwise,} 
	\end{cases}
	\;\;\;\;\;\;\;\;\; \forall \;N.
	\label{eqn:derivation_6}
\end{equation}

\noindent This can be understood as summing along a chosen anti-diagonal line of the joint-photon number distribution (see Fig.~\ref{fig:Inverse_Engineering_Sketch}) corresponding to total photon number $N$ and noting only the axes probabilities are non-zero.  Eq.~\ref{eqn:derivation_6} tells us that the sequence of transformations coinciding with an ANLMZI results in the state

\begin{equation}
	\ket{N,0}_{a,b} \stackrel{\text{ANLMZI}}{\longrightarrow} \frac{e^{i\tfrac{\pi}{4}}}{\sqrt{2}}\left(\ket{N,0}_{a,b} - i\ket{0,N}_{a,b}\right).
	\label{eqn:derivation_7}
\end{equation}

\noindent From this it is easy to show that this will hold for any superposition of Fock states.  Consider the state $\ket{\psi}=\sum_{n}c_n\ket{n}$.  If we describe the ANLMZI as a single operator  $\hat{\mathcal{U}}_{\text{ANLMZI}}$, then this general state transformations according to 

\begin{align}
	\ket{\psi,0}_{a,b}&=\sum_{n=0}^{\infty}c_n\ket{n,0}_{a,b} \stackrel{}{\longrightarrow} \sum_{n=0}^{\infty}c_n \left(\hat{\mathcal{U}}_{\text{ANLMZI}}\ket{n,0}_{a,b}\right), \nonumber \\
	& \longrightarrow \sum_{n=0}^{\infty}c_n \left[\frac{e^{i\tfrac{\pi}{4}}}{\sqrt{2}}\left(\ket{n,0}_{a,b} - i\ket{0,n}_{a,b}\right)\right], \nonumber \\
	& \longrightarrow \frac{e^{i\tfrac{\pi}{4}}}{\sqrt{2}}\left(\ket{\psi,0}_{a,b} - i\ket{0,\psi}_{a,b}\right),
	\label{eqn:derivation_8}
\end{align}
\noindent thus showing that any single-mode state can be mapped to the axes of a two-mode distribution via an ANLMZI.  
Eq.(\ref{eqn:derivation_8}) is one of the main results of this work.

\smallskip
\noindent Next we consider the case where both intermediary modes pass through a cross-Kerr non-linear medium. In this case the cross-Kerr interaction is described by the unitary operation

\begin{equation}
	\hat{\mathcal{U}}^{\left(a,b\right)}_{\text{cross-Kerr}}\left(\kappa\right) = e^{-i\kappa\hat{a}^\dagger\hat{b}^\dagger\hat{a}\hat{b}},
	\label{eqn:derivation_9}
\end{equation}

\noindent in lieu of the intermediary self-Kerr interaction that acts on just the intermediary $a$-mode. Consequently, the phases now work out differently:  The state prior to the second beam splitter is now written as

\begin{equation}
	\ket{\tilde{\psi}_\text{pre-BS2}} =  \sum_{n=0}^{N}C_{n}^{\left(N\right)}\left(-i\right)^{n^2}i^{nN}\ket{n,N-n}_{a,b},
	\label{eqn:derivation_10}
\end{equation}

\noindent where Eq.~\ref{eqn:derivation_4} is recovered after an $N$-dependent phase shift  $\hat{\mathcal{U}}^{\left(a\right)}_{\text{PS}}\left(-\tfrac{N\pi}{2}\right)$.  Since it is experimentally impractical to dynamically change the linear phase shift contingent on the number of photons passing through the interferometer, this interaction will not be suitable for use with continuous-variable states. However, the use of a cross-Kerr interaction remains a viable means of generating $N$-photon $N00N$ states.  

\subsection{\label{subsec:InverseEngineer} Producing N00N state superpositions through inverse-engineering techniques}

\begin{figure}
	\includegraphics[width=0.90\linewidth,keepaspectratio]{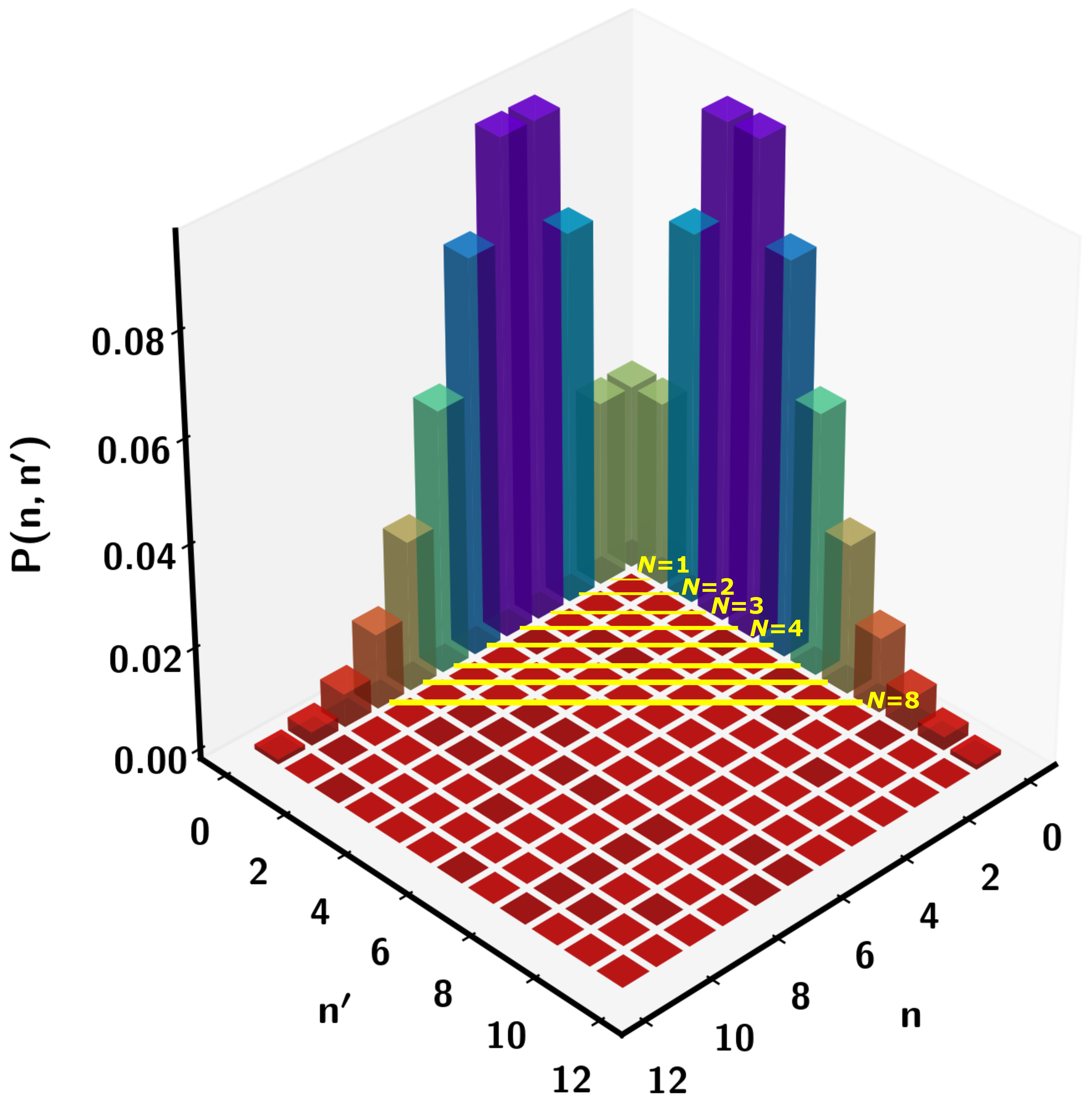}
	\caption{Two-mode joint-photon number distribution for a coherent superpositin of $N00N$ states $\ket{\alpha,0}+\ket{0,\alpha}$, for $|\alpha|^2=4$.  The yellow anti-diagonal lines represent lines of total photon number $N=n+n'$.}
	\label{fig:Inverse_Engineering_Sketch}
\end{figure}

\noindent Working towards the goal of determining possible interactions that can create NOON state superpositions, we will outline a general technique for numerically inverse-engineering such states by working backwards from a beam splitter. We start from the most general of two-mode states 

\begin{equation}
	\ket{\Psi}=\sum_{n=0}^{\infty}\sum_{n'=0}^{\infty}C_{n,n'}\ket{n}_a\ket{n'}_b,
	\label{eqn:inverse_1}
\end{equation}

\noindent and allow the state to interact at a beam splitter such that the final state is given by

\begin{equation}
    \ket{\Psi_F} = \hat{\mathcal{U}}_{\text{BS}}^{(a,b)}\ket{\Psi}=\sum_{n=0}^{\infty}\sum_{n'=0}^{\infty}\tilde{C}_{n,n'}\ket{n,n'}_{a,b}.
    \label{eqn:inverse_1b}
\end{equation}
    
\noindent For each total photon number $N=n+n'$, there is an anti-diagonal along the joint-photon number distribution (see Fig.~\ref{fig:Inverse_Engineering_Sketch}).  As a demonstration, we stipulate that the output state probability amplitudes are pre-determined to be   

\begin{align}
	\tilde{C}_{n,n'} &= \sum_{q=-J}^{J}C_{J+q,J-q}i^{q-J}d_{M,q}^{J}\left(\frac{\pi}{2}\right) \nonumber\\
	&=
	\begin{cases}
		A_N & n=N\;\text{and}\;n'=0 ,\\
		-A_N & n=0\;\text{and}\;n'=N, \\
		0 & \text{otherwise}, \\
	\end{cases}
	\label{eqn:inverse_2}
\end{align}

\begin{figure*}
	\includegraphics[width=0.95\linewidth,keepaspectratio]{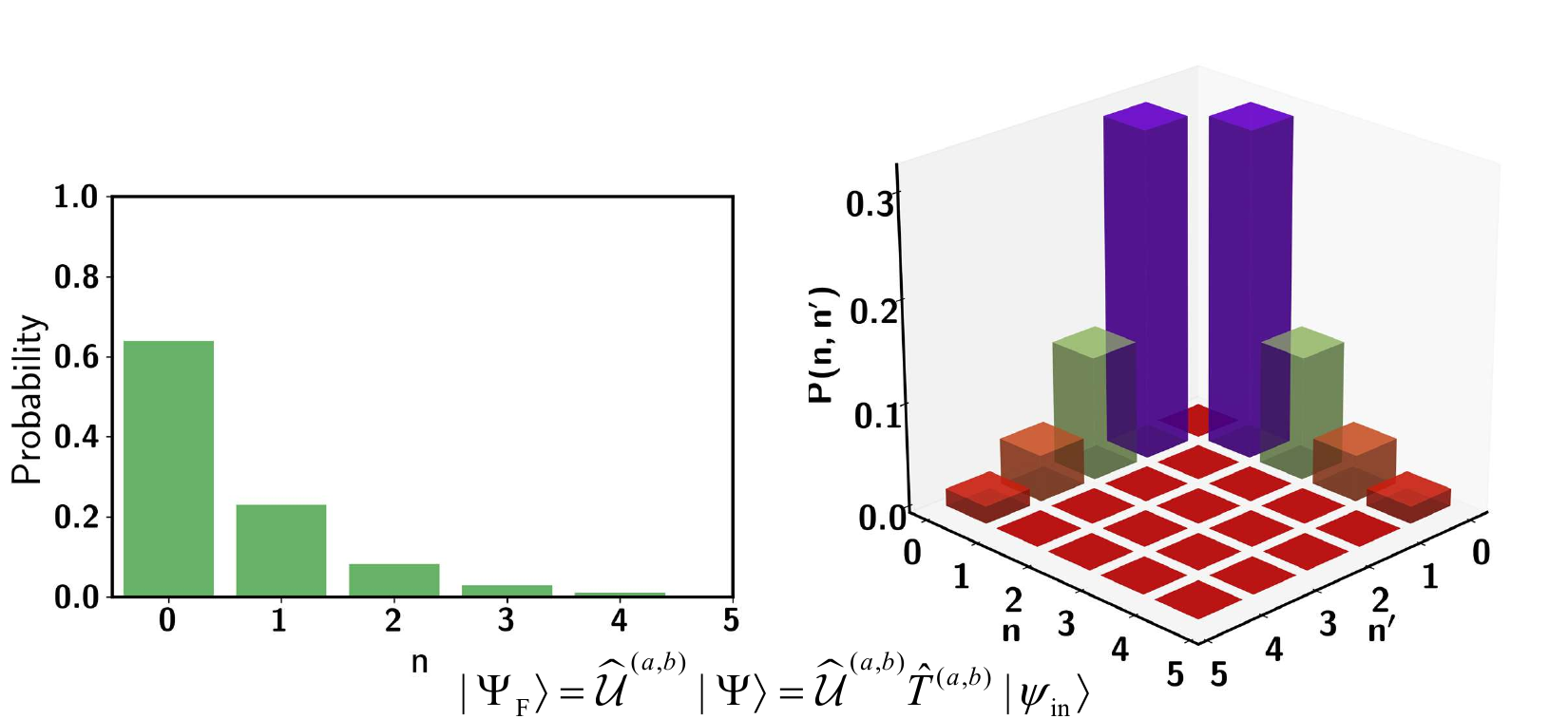}
	\caption{A demonstration using the $\hat{T}^{(a,b)}$ matrix of Eq.~\ref{eqn:inverse_5} to map a single-mode pure ``thermal" state $\ket{z}$ into a superposition of $N00N$ states of the form $\propto\big(\ket{z,0}_{a,b}-\ket{0,z}_{a,b}\big)$.}
	\label{fig:mapping_example}
\end{figure*}

\noindent where $J=\tfrac{1}{2}\left(n+n'\right)$ and $M=\tfrac{1}{2}\left(n-n'\right)$.  Specifically, we presuppose that the distribution is populated only the axes and differ only by a $\pi$-phase shift.  This results in a system of linear equations for each value of $N$ (each anti-diagonal) which can be solved to express the initial state coefficients in terms of the axes probabilities $\left\{A_N\right\}$. For the case of $n,n'\in \left[0..4\right]$ the input state coefficients will be 

\begin{align}
	&C_{n,n'} = \nonumber\\
		&
		\begin{pmatrix}
			0 & e^{i\frac{\pi}{4}}A_1 & A_2 & \frac{e^{-i\frac{\pi}{4}}}{\sqrt{4}}A_3 & 0 \\
			-e^{i\frac{\pi}{4}}A_1 & 0 & \frac{\sqrt{3}e^{-i\frac{\pi}{4}}}{\sqrt{4}}A_3 & -iA_4 & 0 \\
			-A_2 & -\frac{\sqrt{3}e^{-i\frac{\pi}{4}}}{\sqrt{4}}A_3 & 0 & 0 & 0 \\
			-\frac{e^{-i\frac{\pi}{4}}}{\sqrt{4}}A_3 & iA_4 & 0 & 0 & 0 \\
			0 & 0 & 0 & 0 & 0 
	\end{pmatrix}.
	\label{eqn:inverse_3}
\end{align}

\noindent Interestingly for this example, the initial state cannot contain any correlated states of the form $\ket{n,n}_{a,b}$. As discussed in Section \ref{sec:interferometry}, such states can be initially produced via mixing (non-classical) states of definite odd parity with any other state at a beam splitter \cite{ref:Alsing_eHOM}. We can then map backwards from the supposition that the input state is of the form

\begin{equation}
	\ket{\psi_\text{in}} = \ket{\psi}_a\otimes\ket{0}_b=
		\begin{pmatrix}
			A_0  \\
			A_1 \\
			A_2 \\
			\vdots
		\end{pmatrix}
	\otimes
		\begin{pmatrix}
			1  \\
			0 \\
			0 \\
			\vdots
		\end{pmatrix} =
		\begin{pmatrix}
			A_0  \\
			0    \\
			\vdots \\
			A_1\\
			\vdots
		\end{pmatrix},	
	\label{eqn:inverse_4}
\end{equation}   

\noindent to find the transformation $\hat{T}^{(a,b)}$ such that

\begin{equation}
    \ket{\Psi}=\hat{T}^{(a,b)}\ket{\psi_\text{in}}.
    \label{eqn:inverse_4b}
\end{equation}

\noindent The $25\times25$ matrix $\hat{T}^{(a,b)}$ that transforms the state of Eq.~\ref{eqn:inverse_4} to the two-mode state with coefficients given by Eq.~\ref{eqn:inverse_3} is found to have the form of an upper-right triangular matrix with non-zero elements (where we drop the superscript from this point on for notational convenience)

\begin{align}
	\hat{T}_{5,2} &= -\hat{T}_{5,5} = e^{i\tfrac{\pi}{4}},\;&&\hat{T}_{10,3} = -\hat{T}_{10,10} = 1,\nonumber\\
	\hat{T}_{15,3} &= -\hat{T}_{15,15} = \tfrac{1}{\sqrt{4}}e^{i\tfrac{\pi}{4}},\;&&\hat{T}_{15,7} = -\hat{T}_{15,11} = \sqrt{\tfrac{3}{4}}e^{-i\tfrac{\pi}{4}},\nonumber\\
	\hat{T}_{20,8} &= -\hat{T}_{20,16} = -i.
	\label{eqn:inverse_5}
\end{align}

\noindent Interestingly, for the case of a $25\times25$ matrix comprised of $625$ elements corresponding to a space size of $5\otimes 5$, only ten elements are responsible for generating the input state of the form Eq.~\ref{eqn:inverse_3}. Due to the structure of the initial state Eq.~\ref{eqn:inverse_4}, many of the zero-valued elements can take on any other value without affecting the result of the transformation.  For this reason, the resulting matrix using the elements of Eq.~\ref{eqn:inverse_5} only constitutes a particular solution.  Furthermore, one can form the matrix using the elements of Eq.~\ref{eqn:inverse_5} to find

\begin{equation}
	\hat{T}^{(a,b)\dagger}\hat{T}^{(a,b)} = 2\left(\hat{I}_a-\hat{P}_0^{\left(a\right)}\right)\otimes\hat{P}_0^{\left(b\right)},
	\label{eqn:inverse_7}
\end{equation}

\noindent where $\hat{I},\hat{P}_0$ are the identity and zero-photon projection operators, respectively.  The state normalization factor is then given by $\braket{\hat{T}^{(a,b)\dagger}\hat{T}^{(a,b)}}^{-1/2}$. We plot an example of a mapping using this transformation matrix in Fig.~\ref{fig:mapping_example} for the case of an initial state $\ket{\psi_\text{in}}=\ket{z}_a\ket{0}_b$, where

\begin{equation}
    \ket{z}=\big(1-|z|^2\big)^{1/2}\sum_{n=0}^\infty z^n\ket{n}_a
    \label{eqn:inverse_5a}
\end{equation}

\noindent is a fictitious state representing a single-mode pure state with thermal-state statistics.  Note that $\hat{T}^{(a,b)}$ does not represent a unitary transformation. However, judging by its form a reasonable assumption for $\hat{T}^{(a,b)}$ is that it could be expressed generally, for example, as a sum of boson-mode operations of the form

\begin{equation}
	\hat{T}^{(a,b)} \stackrel{?}{\simeq} \text{Exp}\left[\sum_i\sum_j e^{i\beta_{i,j}\hat{a}^{i}\hat{b}^j} + e^{\gamma_{i,j}\hat{a}^{\dagger\;i}\hat{b}^{\dagger\;j}}\right],
	\label{eqn:inverse_6}
\end{equation} 

\noindent where $\beta_{i,j},\gamma_{i,j}$ can be complex.  Such an operation could, in practice, correspond to a sequence of photon-subtractions, additions or some other form of state-reductive measurement. As an example, consider the initial state $\ket{\psi_{\text{in}}}=\ket{1,0}$ and a $\hat{J}_y$ beam splitter such that

\begin{equation}
    \hat{\mathcal{U}}_{\text{BS}}^{(a,b)}\ket{0,1}_{a,b} = e^{i\tfrac{\pi}{2}\hat{J}_y}\ket{0,1}_{a,b} = \frac{1}{\sqrt{2}}\left( \ket{0,1}-\ket{1,0} \right).
    \label{eqn:inverse_6b}
\end{equation}

\noindent We can set up the system of linear equations similarly to Eq.~\ref{eqn:inverse_2} where $A_n=\delta_{n,1}$ to find the inverse-mapping transformation  

\begin{equation}
    \hat{T}^{(a,b)} = \sqrt{2}
            \begin{pmatrix}
			0 & 0 & 0 & 0\\
			0 & 0 & 1 & 0\\
			0 & 0 & 0 & 0\\
			0 & 0 & 0 & 0
		\end{pmatrix}
    \propto \hat{a}\hat{b}^\dagger
    \label{eqn:inverse_6c}
\end{equation}

\noindent which yields the transformation

\begin{equation}
    \ket{\Psi} = \hat{T}^{(a,b)}\ket{\psi_{\text{in}}} \propto \hat{a}\hat{b}^\dagger\ket{1,0}_{a,b}=\ket{0,1}_{a,b},
    \label{eqn:inverse_6d}
\end{equation}

\noindent where the normalization factor can be found with respect to the initial state $\ket{\psi_{\text{in}}}$ to be $\braket{\hat{T}^{(a,b)\dagger}\hat{T}^{(a,b)}}^{-1/2}=1/\sqrt{2}$.  Taking the occupation number in the initial state one higher and assuming the same beam splitter type, we have that for the input state $\ket{\psi_{\text{in}}}=\ket{2,0}$ mapping to the output state $\ket{\Psi_{F}}\propto\left(\ket{0,2}-\ket{2,0}\right)$, the intermediate state will be given by $\ket{\Psi}\equiv\ket{1,1}$ as a consequence of the Hong-Ou-Mandel effect.  This dictates that the transformation that maps $\ket{\psi_{\text{in}}}\to\ket{\Psi}$ will again be, by inspection, proportional to $\hat{a}\hat{b}^\dagger$. Using the techniques outlined above, a particular solution for $\hat{T}^{(a,b)}$ can be found numerically for this case, which can be written as (dropping subscripts)

\begin{align}
    \hat{T}^{(a,b)} &= \hat{a}\hat{b}^\dagger + \big(\sqrt{2}-1\big)\ket{0,1}\bra{1,0} - \sqrt{2}\ket{0,2}\bra{1,1}- \nonumber\\
    & \;\;\;\;\;\;\;\;\;\;\;\;\;\;\;\;\;\;\;\;\;\;\;\;\;\;\;\;\;\;\;\;\;\;\;\;\;\;\;\;\;\;\;\;\;\;\;\; - 2\ket{1,2}\bra{2,1}.
	\label{eqn:inverse_6e}
\end{align}

\noindent Notice for the initial state $\ket{\psi_\text{in}}=\ket{2,0}$, all of the operators beyond $\hat{a}\hat{b}^\dagger$ in Eq.~\ref{eqn:inverse_6e} will vanish. From this, and recalling that the operator $\hat{T}^{(a,b)}$ will map \textit{any} initial state of the form in Eq.~\ref{eqn:inverse_4} to a state that will produce a $N00N$ state superposition upon beam splitting, one can conclude that for the particular case of $\ket{\psi_\text{in}}=\ket{2,0}$, the transformation $\hat{T}^{(a,b)}$ is realized by $\hat{a}\hat{b}^\dagger$. Thus, while $\hat{T}^{(a,b)}$ represents a general transformation, its utility is dependent on the elements of the initial state, and may be realized via boson mode operations such as photon addition and/or subtraction. Further to that point, photon subtraction can be realized experimentally by mixing one mode at a high-transmittance beam splitter with a vacuum state and heralding off a particular photon-number detection corresponding to the number of photons being subtracted.  \\

\noindent As a more concrete demonstration, it is not hard to 
engineer
the output coefficients in Eq.~\ref{eqn:inverse_2} 
to be 
of the form 
\begin{equation}
	\tilde{C}_{n,n'} =e^{i\tfrac{\pi}{4}} \times
	\begin{cases}
		i^NA_N & n=N\;\text{and}\;n'=0 ,\\
		\left(-1\right)^{N}A_N & n=0\;\;\text{and}\;n'=N, \\
		0 & \text{otherwise}. \\
	\end{cases}
	\label{eqn:inverse_8}
\end{equation}

\noindent The transformation required to generate such a superposition state can be determined using the methods discussed above. We find that this transformation is realized by the sequence of unitary transformations

\begin{equation}
	\hat{T}^{(a,b)} \equiv \mathcal{\hat{U}}^{\left(a\right)}_{\text{PS}}\left(\tfrac{\pi}{2}\right)\mathcal{\hat{U}}^{\left(a\right)}_{\text{self-Kerr}}\left(\tfrac{\pi}{2}\right)\mathcal{\hat{U}}^{\left(a,b\right)}_{\text{BS}}\left(\tfrac{\pi}{2}\right),
	\label{eqn:inverse_9}
\end{equation}

\noindent which produces, up to the post-BS phase-shifters, the state of equation Eq.~\ref{eqn:derivation_8}. One can conclude from this that for the case of a $N00N$ state superposition where the normalization is not dependent on the single-mode state coefficients, the $\hat{T}$ transformation can be found to be unitary, as is the case for coefficients of the form Eq.~\ref{eqn:inverse_8}.  For the case of Eq.~\ref{eqn:inverse_2} however, the state normalization is dependent on the single-mode state zero-photon probability $|A_{0}|^2$.  A valid strategy would be to start from the premise of a superposition of normalized $N00N$ states at the output of a balanced beam splitter and work backwards to find the unitary transformations that produce them.  More generally, one can consider the case where the state probability amplitudes at the output of a balanced beam splitter, as per Eq.~\ref{eqn:inverse_2}, are given by  

\begin{equation}
	\tilde{C}_{n,n'} = 
	\begin{cases}
		e^{i\lambda_N^{(a)}}A_N & n=N\;\text{and}\;n'=0 ,\\
		e^{i\lambda^{(b)}_N}A_N & n=0\;\;\text{and}\;n'=N, \\
		0 & \text{otherwise}, \\
	\end{cases}
	\label{eqn:inverse_10}
\end{equation}

\noindent and one endeavours to determine the form of $\hat{T}^{(a,b)}$ to map the state $\ket{\psi,0}_{a,b}=\sum_{n=0}^{\infty}A_n\ket{n,0}_{a,b}$ to the output state with coefficients given by Eqn.~\ref{eqn:inverse_10}.  The investigation of these techniques to probe the generation of $N00N$ state superpositions and other states of desirable properties remains an ongoing subject of research.   

\section{\label{sec:closing} Conclusion}

\noindent $N00N$ states and their superpositions have long been discussed in the context of quantum optical interferometry as they have been shown to yield Heisenberg-limited phase sensitivity when paired with the ideal detection observable (i.e. photon-number parity-based measurements). In this work, we have demonstrated a means of producing generalized superpositions of N00N states weighted on each of the axes of the two-mode joint-photon number distribution by the statistics of any single-mode pure state. Our scheme requires the use of an asymmetric non-linear MZI characterized by a self-Kerr interaction on one of the intermediary modes of the interferometer. We further showed how one can generate the $N$ photon $N00N$ state through the use of a non-linear MZI characterized by a cross-Kerr interaction between the two intermediary modes of the MZI.  We note that this would require an $N$-dependent phase shift on one of the modes, making it suitable for an $N$-photon Fock state (while unsuitable for a general superposition state). \\

\noindent Additionally, and within the context of generating $N00N$ state superpositions, we reviewed an extension of the HOM effect for which a non-classical input state of definite odd parity displays a contiguous line of zero alone the $n_a=n_b$ diagonal of the output distribution (number-resolved coincident detections) when mixed with any other state at a beam splitter. This is referred to as the central nodal line (CNL). For non-classical inputs of definite even or odd parity, the resulting two-mode distribution displays non-diagonal sequences of bifurcations designated as pseudo-nodal curves (PNCs).  One feature of this non-classical interference effect that can be observed is the migration of the peak probabilities towards the axes of the joint-photon number distribution, reminiscent of the well-known $N00N$ state. With this as motivation, we introduce inverse-engineering techniques to probe the means for mapping single-mode states into superpositions of $N00N$ states. We note that while we introduce these techniques in the context of generating $N00N$ state superpositions, the techniques themselves are general and can be used to generate symmetric states of any desirable quantum properties.\\

\section{\label{sec:acknowledge} Acknowledgments}

\noindent The authors wish to dedicate this work to the memory of Jonathan P. Dowling, a friend, mentor to many, and a truly unique, dynamic colleague, who made seminal contributions to the field of quantum optics, and in particular, the study, generation, and applications of $N00N$ states to quantum information science and technlogy.
\smallskip

\noindent RJB would like to thank the Griffiss Institute (GI) for support of this work. PMA, JS and CCG would like to acknowledge support from the Air Force Office of Scientific Research (AFOSR). Any opinions, findings and conclusions or recommendations expressed in this material are those of the author(s) and do not necessarily reflect the views of Air Force Research Laboratory or the U.S. Navy.

\section{\label{sec:declarations} Author Declarations}

\subsection{\label{subsec:conflicts} Conflict of interest}

\noindent The authors have no conflicts to disclose.

\section{\label{sec:data} Data Availability}

\noindent The data that support the findings of this study are available from the corresponding author upon reasonable request.

\section{\label{sec:appendicies} Appendices}
\subsection{\label{app:secA} The Schwinger realization of the SU(2) Lie algebra}

\noindent Here we will provide a brief review of the Schwinger representation of the SU(2) Lie algebra.  For a more comprehensive discussion on the topic, see for example Yurke \textit{et al.} \cite{ref:Yurke_SU2} or Birrittella \textit{et al.} \cite{ref:Birrittella_ParityReview}.  Consider a two mode field with creation and annihilation operators satisfying the usual boson commutation relations $\left[\hat{a}_{i},\hat{a}_{j}\right] = \big[\hat{a}^{\dagger}_{i},\hat{a}^{\dagger}_{j}\big] = 0$ and $\big[\hat{a}_{i},\hat{a}^{\dagger}_{j}\big] = \delta_{i,j}$. One can introduce the Hermitian operators

\begin{align}
	\hat{J}_{x} &= \dfrac{1}{2}\big(\hat{a}_{1}^{\dagger}\hat{a}_{2} + \hat{a}^{\dagger}_{2}\hat{a}_{1}\big), \nonumber \\
	\hat{J}_{y} &= -\dfrac{i}{2}\big(\hat{a}_{1}^{\dagger}\hat{a}_{2} - \hat{a}^{\dagger}_{2}\hat{a}_{1}\big), \label{eqn:Aa1}\\
	\hat{J}_{z} &= \dfrac{1}{2}\big(\hat{a}_{1}^{\dagger}\hat{a}_{1} - \hat{a}^{\dagger}_{2}\hat{a}_{2}\big), \nonumber
\end{align}

\noindent and $\hat{N} = \big(\hat{a}_{1}^{\dagger}\hat{a}_{1} + \hat{a}^{\dagger}_{2}\hat{a}_{2}\big)$, satisfying the commutation relations of the Lie algebra of SU(2):

\begin{equation}
	\big[\hat{J}_{i},\hat{J}_{j}\big] = i\hat{J}_{k}\epsilon_{i,j,k}.
	\label{eqn:Aa2}
\end{equation}

\noindent Note that the operator $\hat{N}$ commutes will all operators in Eq.~\ref{eqn:Aa1}.  One can also define the operator $\hat{J}_{0}=\tfrac{1}{2}\hat{N}$ such that $\hat{J}_{0}\ket{j,m}=j\ket{j,m}$. The Casimir invariant for the group is then given by $\hat{J}^{2} = \hat{J}_{x}^{2} + \hat{J}_{y}^{2} + \hat{J}_{z}^{2}=\hat{J}_0\left(\hat{J}_0 + 1\right)$.  It is also useful to recall the action of the angular momentum operatos $\hat{J}_{i}$ on the states $\ket{j,m}$:

\begin{align}
	\hat{J}^{2}\ket{j,m} &= j\left(j+1\right)\ket{j,m} \nonumber \\
	\hat{J}_{z}\ket{j,m} &= m\ket{j,m} \\
	\hat{J}_{\pm}\ket{j,m} &= \sqrt{j\left(j+1\right) -m\left(m\pm 1\right)}\ket{j,m\pm 1}, \nonumber 
	\label{eqn:Aa2.1}
\end{align}

\noindent where the ladder operators are given by 

\begin{equation}
	\hat{J}_{\pm} = \hat{J}_{x} \pm i\hat{J}_{y}.
	\label{eqn:Aa2.1.1}
\end{equation}

\noindent For bosons, we can freely change representation between the `angular momentum' basis with states $\ket{j,m}$ and two-mode Fock basis with states $\ket{n}_a\otimes\ket{n'}_b$.\\

\noindent A beam splitter transforms the input mode boson operators according to the scattering matrix of the device, that is

\begin{equation}
	\vec{\hat{a}}_{\text{out}} = \hat{U}\;\vec{\hat{a}}_{\text{in}}
	\;\;\;\to\;\;\;
	\begin{pmatrix}
		\hat{a}_{1}  \\
		\hat{a}_{2} 
	\end{pmatrix}_{\text{out}} 
	=
	\begin{pmatrix}
		U_{11} & U_{12} \\
		U_{21} & U_{22}
	\end{pmatrix} 
	\begin{pmatrix}
		\hat{a}_{1}  \\
		\hat{a}_{2} 
	\end{pmatrix}_{\text{in}}.
	\label{eqn:Aa3}
\end{equation}

\noindent Note that since the boson creation and annihilation operators must satisfy the commutation relations both before and after beam splitting, the matrix $\hat{U}$ must be unitary. We will briefly show how this transforms the operators of SU(2), $\vec{J} = \big(\hat{J}_{x},\hat{J}_{y},\hat{J}_{z}\big)$.  Consider the scattering matrix $\hat{U}$ of the form

\begin{equation}
	\hat{U} = 
	\begin{pmatrix}
		\cos\tfrac{\theta}{2} & -i\sin\tfrac{\theta}{2} \\
		-i\sin\tfrac{\theta}{2} & \cos\tfrac{\theta}{2}
	\end{pmatrix} 
	=
	\begin{pmatrix}
		t & -ir \\
		-ir & t
	\end{pmatrix} ,
	\label{eqn:Aa4}
\end{equation}

\noindent which corresponds to a beam splitter with transmittance and reflectivity $T=\cos^{2}\tfrac{\theta}{2}$ and $R=\sin^{2}\tfrac{\theta}{2}$, respectively.  For this scattering matrix, $\vec{J}$ transforms to 

\begin{align} 
	\begin{pmatrix}
		\hat{J}_{x}  \\
		\hat{J}_{y}  \\
		\hat{J}_{z}
	\end{pmatrix}_{\text{out}}
	& =
	\begin{pmatrix}
		1 & 0 & 0  \\
		0 & \cos\theta & -\sin\theta  \\
		0 & \sin\theta & \cos\theta 
	\end{pmatrix}
	\begin{pmatrix}
		\hat{J}_{x}  \\
		\hat{J}_{y}  \\
		\hat{J}_{z}
	\end{pmatrix}_{\text{in}}\nonumber\\
	& = 
	e^{i\theta\hat{J}_{x}}
	\begin{pmatrix}
		\hat{J}_{x}  \\
		\hat{J}_{y}  \\
		\hat{J}_{z}
	\end{pmatrix}_{\text{in}}
	e^{-i\theta\hat{J}_{x}},
	\label{eqn:Aa5}
\end{align}

\noindent which amounts to a rotation about the fictitious $x$-axis.  Note that the last line of Eq.~\ref{eqn:Aa5} can be verified via the use of the Baker-Hausdorff identity

\begin{equation}
	e^{\tau\hat{A}}\hat{B}e^{-\tau\hat{A}} = \hat{B} + \tau\big[\hat{A},\hat{B}\big] + \frac{1}{2}\tau^{2}\big[\hat{A},\big[\hat{A},\hat{B}\big]\big] + ..\;.
	\label{eqn:Aa6}
\end{equation}

\noindent Working in the Schr\"{o}dinger picture, the action of the beam splitter corresponds to a transformation of the initial state given by  

\begin{equation}
	\ket{\psi_\text{out, BS}} = e^{-i\theta\hat{J}_{x}}\ket{\psi_\text{in}}.
	\label{eqn:Aa7}
\end{equation}

\noindent We can also express a two-mode state in the Fock basis in terms of the basis states of SU(2) (angular momentum states) using Eq.~\ref{eqn:Aa1}, yielding (dropping subscripts for notational convenience)

\begin{align}
	&\hat{J}_{z}\ket{j,m} = m\ket{j,m} \;\;\leftrightarrow\;\; \hat{J}_{z}\ket{n,n'} = \frac{n-n'}{2}\ket{n,n'}, \nonumber \\
	\label{eqn:Aa8}\\
	&\hat{J}^{2}\ket{j,m}=j\big(j+1\big)\ket{j,m}\;\leftrightarrow\;  \nonumber\\
	&\;\;\;\;\;\;\;\;\;\;\;\;\;\;\;\;\;\;\;\;\;\hat{J}^{2}\ket{n,n'} = \frac{n+n'}{2}\left(\frac{n+n'}{2} + 1\right)\ket{n,n'},
\end{align}

\noindent which informs $\ket{n,n'}_{a,b}\to\ket{j,m}$ where the values of $j$ and $m$ are given by $j=\tfrac{n+n'}{2}$ and $m=\tfrac{n-n'}{2}$. Inversely $\ket{j,m}\to\ket{n,n'}_{a,b}$ where $n=j+m$ and $n'=j-m$ with $n+n'=2j$ and $m \in \{-j,...,j\}$. With this, the connection between two-mode boson fields and the `angular momentum' states of SU(2) is now complete.  

\subsection{\label{app:secB}Elements of the Wigner-$d$ rotation matrix}

\noindent Here we provide a brief discussion of the matrix elements of an arbitrary rotation specified by an axis of rotation $\boldsymbol{\hat{n}}$ and angle of rotation $\phi$.  The matrix elements, with $\hbar \to 1$ for convenience, are 

\begin{equation}
	\mathcal{D}_{m',m}^{j}\left(R\right) = \braket{j,m'|e^{-i\phi\; \boldsymbol{J}\cdot\boldsymbol{\hat{n}}}|j,m}.
	\label{eqn:Ab1}
\end{equation}

\noindent Since the rotation operator commutes with the $\hat{J}^{2}$ operator, a rotation cannot change the $j$ value of a state. The $\big(2j+1\big)\times\big(2j+1\big)$ matrix formed by $\mathcal{D}_{m',m}^{j}\big(R\big)$ is referred to as the $\big(2j+1\big)$-dimensional irreducible representation of the rotation operator $\mathcal{D}\big(R\big)$ .  We now consider the matrix realization of the Euler Rotation,

\begin{align}
	\mathcal{D}_{m',m}^{j}\left(\alpha,\beta,\gamma\right) &= \braket{j,m'|R_{z_{f}}\left(\alpha\right)R_{y_{f}}\left(\beta\right)R_{z_{f}}\left(\gamma\right)|j,m} \nonumber\\
	&= \braket{j,m'|e^{-i\alpha\hat{J}_{z}}e^{-i\beta\hat{J}_{y}}e^{-i\gamma\hat{J}_{z}}|j,m}.
	\label{eqn:Ab2}
\end{align} 

\noindent These matrix elements are referred to as the Wigner-$D$ rotation elements. Notice that the first and last rotation only add a phase factor to the expression, thus making only the rotation about the fixed $y$-axis the only non-trival part of the matrix.  For this reason, the Wigner-$D$ matrix elements are written in terms of a new matrix

\begin{align}
	\mathcal{D}_{m',m}^{j}\left(\alpha,\beta,\gamma\right) &=\braket{j,m'|e^{-i\alpha\hat{J}_{z}}e^{-i\beta\hat{J}_{y}}e^{-i\gamma\hat{J}_{z}}|j,m} \nonumber \\
	&= e^{-i\left(m'\alpha + m\gamma\right)}\braket{j,m'|e^{-i\beta\hat{J}_{y}}|j,m} \nonumber\\
	&= e^{-i\left(m'\alpha + m\gamma\right)}\; d_{m',m}^{j}\left(\beta\right),
	\label{eqn:Ab3}
\end{align}

\noindent where the matrix elements $d_{m',m}^{j}\left(\beta\right) = \braket{j,m'|e^{-i\beta\hat{J}_{y}}|j,m}$ are formally known as the Wigner-$d$ rotation elements and are given by

\begin{align}
	d_{m',m}^{j}&\left(\beta\right) = \Bigg(  \dfrac{\big(j-m\big)!\big(j+m'\big)!}{\big(j+m\big)!\big(j-m'\big)!}  \Bigg)^{1/2} \times\nonumber \\
	&\times\dfrac{\big(-1\big)^{m'-m}\cos^{2j+m-m'}\big(\tfrac{\beta}{2}\big)\sin^{m'-m}\big(\tfrac{\beta}{2}\big)}{\big(m'-m\big)!}\times \nonumber \\
	&\times \;_{2}F_{1}\left(m'-j,-m-j;m'-m+1;-\tan^{2}\big(\tfrac{\beta}{2}\big)\right),
	\label{eqn:Ab4}
\end{align}

\noindent with the property

\begin{equation}
	d_{m'm}^{j}\left(\beta\right) = 
	\begin{cases}
		d_{m',m}^{j}\left(\beta\right) & m' \geq m \\
		\\
		d_{m,m'}^{j}\left(-\beta\right) & m' < m 
	\end{cases}
	\label{eqn:Ab5}
\end{equation}

\noindent and where $_{2}F_{1}\big(a,b;c;z\big)$ is a hypergeometric function.  It is worth noting that in typical interferometric calculations, one naturally ends up with an expression that depends on the Wigner-$d$ matrix elements.  However, when simply dealing with a single $\hat{J}_{x}$-type beam splitter of angle $\theta$, one encounters the matrix elements $\braket{j,m'|e^{-i\theta\hat{J}_{x}}|j,m}$.  This can be simplified using the Baker-Hausdorff identity of Eq.~\ref{eqn:Aa6} to

\begin{align}
	\braket{j,m'|e^{-i\theta\hat{J}_{x}}|j,m} &= \braket{j,m'|e^{i\tfrac{\pi}{2}\hat{J}_{z}}e^{-i\theta\hat{J}_{y}}e^{-i\tfrac{\pi}{2}\hat{J}_{z}}|j,m} \nonumber\\
	&= \mathcal{D}_{m',m}^{j}\left(-\tfrac{\pi}{2},\theta,\tfrac{\pi}{2}\right)\nonumber \\
	&= i^{m'-m}d_{m',m}^{j}\left(\theta\right).
	\label{eqn:Ab7}
\end{align}

\noindent Lastly, the beam splitter coefficients $f^{(n,m)}_p(\theta)$ \cite{ref:Agarwal_textbook} 
discussed in the extended HOM effect,  Alsing \textit{et al.} \cite{ref:Alsing_eHOM}, 
are related to the 
the Wigner rotation matrices $d^j_{m',m}(\theta)$ employed in this work
via
\begin{equation}
 f^{(n,m)}_p(\theta) \equiv d^{(n+m)/2}_{p-(n+m)/2, (n-m)/2}(\theta),
 \label{eqn:Ab8}
\end{equation}
where the output state of the dual-mode Fock input state $|n,m\rangle_{\text{BS-in}}$ to a beam splitter is given by
$\ket{n,m}_{\text{BS-out}}=\sum_{p=0}^{n+m}\,f^{(n,m)}_p(\theta)\,|p,n+m-p\rangle$.
\\

\noindent For a more comprehensive analysis of these matrix elements for high-spin numerical evaluation \cite{ref:Tajima}, as well as a detailed list of properties see Tajima \textit{et al.} \cite{ref:Tajima} and Birrittella \textit{et al.} \cite{ref:Birrittella_ParityReview}.

\nocite{*}
\bibliography{N00N_Map_Bib}

\end{document}